\documentclass[twocolumn ]{aastex61}

\newcommand\aastex{AAS\TeX}

\newcommand{\FeXXIV}{\hbox{Fe {\sc xxiv}}}
\newcommand{\FeXXV}{\hbox{Fe {\sc xxv}}}
\newcommand{\FeXXVI}{\hbox{Fe {\sc xxvi}}}
\newcommand{\SiXIV}{\hbox{Si {\sc xiv}}}

\newcommand{\SXVI}{\hbox{S {\sc xvi}}}
\newcommand{\NeX}{\hbox{Ne {\sc x}}}

\newcommand{\simgt}{\lower 2pt \hbox{$\, \buildrel {\scriptstyle >}\over {\scriptstyle\sim}\,$}}
\newcommand{\simlt}{\lower 2pt \hbox{$\, \buildrel {\scriptstyle <}\over {\scriptstyle\sim}\,$}}

\newcommand{\iras}{IRAS~13224$-$3809}
\newcommand{\xmm}{\emph{XMM-Newton}}

\received{June 11, 2018}
\accepted{September 22, 2018}
\submitjournal{ApJ}

\shorttitle{\aastex\ sample article}
\shorttitle{The Variable Outflow of \iras}
\shortauthors{Chartas \& Canas}

\begin{document}

\def\sarc{$^{\prime\prime}\!\!.$}

\title{The Variable Relativistic Outflow of \iras}

\correspondingauthor{George Chartas}
\email{chartasg@cofc.edu}

\author[0000-0003-1697-6596]{George Chartas}
\affil{Department of Physics and Astronomy \\
College of Charleston \\
Charleston, SC, 29424, USA}

\author{Manuel H. Canas}
\affil{Department of Physics and Astronomy \\
College of Charleston \\
Charleston, SC, 29424, USA}








\begin{abstract}
The discovery of an ultrafast outflow has been reported in the $z$=0.0658 narrow line Seyfert galaxy \iras\ \citep{2017MNRAS.469.1553P}.
The ultrafast outflow was first inferred through the detection of highly blueshifted absorption lines \citep{2017MNRAS.469.1553P} and then confirmed with a principal component analysis (PCA) \citep{2017Natur.543...83P}. Two of the reported properties of this outflow differed from those typically detected in other AGN with ultrafast outflows.
First, the outflow velocity was found not to vary with $v$=0.236$c$~$\pm$~0.006$c$.
Second, the equivalent width of the highly blueshifted absorption line was reported to be anti-correlated with the 3$-$10~keV flux of this source.
We present a re-analysis of the \xmm\ observations of \iras\ considering the influence of background. We also undertook a different analysis approach in combining the spectra and investigated the change of the properties of the outflow as a function of 3$-$10~keV flux and time. 
We confirm the presence of an ultrafast outflow in \iras, however, we find that the background spectra used in the Parker et al. analyses dominate the source spectra for energies near the blueshifted iron lines.   
By reducing the source extraction regions to improve the signal-to-noise ratio we discover larger than previously reported outflow velocities
and find that the outflow velocity varies from $\sim$~0.2$c$ to $\sim$~0.3$c$ and increases with 3$-$10~keV flux. The previously reported anti-correlation between equivalent width of the iron line and 3$-$10~keV flux disappears when the background spectra are reduced by optimizing the source extraction regions.

\end{abstract}

\keywords{galaxies: formation --- galaxies: evolution --- Seyfert: absorption lines ---X-rays: galaxies ---intergalactic medium} 




\section{Introduction} \label{sec:intro}

Relativistic wide-angle outflows of AGN our now considered one of the main mechanisms regulating the evolution of galaxies through a feedback process
(e.g., see review by \cite{2015ARA&A..53..115K} and references within).   
These wide-angle winds are thought to transfer a substantial amount of their kinetic energy to the surrounding gas resulting in quenching of star formation in the host galaxy  
by heating the interstellar medium or by ejecting the gas from the galaxy (e.g., \cite{2012MNRAS.425..605F}, \cite{2012ApJ...745L..34Z}).
Several of these models consider two main phases of the interaction between the wind and the ISM. In the momentum-conserving phase the relativistic wind collides with the ISM producing a forward and reverse shocked wind that loses a significant amount of its kinetic energy through inverse Compton cooling.  In the energy-conserving phase the wind expands adiabatically and reaches a terminal velocity of a few 1,000 km s$^{-1}$.  
The first evidence of powerful relativistic winds in quasars came from observations of APM 08279$+$5255, PG~1211+143, and PDS~456 \citep{2002ApJ...579..169C,2003MNRAS.345..705P,2003ApJ...593L..65R}.  
Follow-up studies of a larger sample of nearby Seyfert galaxies showed that about 40\% of these AGN have highly-ionized ultrafast outflows (UFOs)
with velocities exceeding 10,000 km~s$^{-1}$ and with average velocities ranging between 0.1$c$ and 0.3$c$ \citep{2010A&A...521A..57T,2013MNRAS.430...60G}.
There have been attempts to compare the energetics of small scale ultrafast outflows with larger scale molecular outflows in galaxies to test feedback models.
The presence of both small and large scale energy-conserving outflows were recently discovered by \cite{2015Natur.519..436T} in the $z$~=~0.189 ULIRG IRAS F11119+3257, 
and by \cite{2015A&A...583A..99F} in the $z = 0.04217$ ULIRG Mrk~231. \cite{2017A&A...608A..30F} have also reported the detection of molecular gas outflowing with maximum velocity of $v$~=~1340~km~s$^{-1}$ in the UFO/BAL quasar APM~08279+5255 at $z$ = 3.912.

Recently, \cite{2017MNRAS.469.1553P} reported the discovery of a relativistic wind in the $z$ = 0.0658 narrow line Seyfert galaxy \iras. The discovery was based on the analysis of \xmm\ observations of \iras\ made in 2016.
These observations had a total exposure time of $\sim$~1.5~Ms spread over a month from 2016$-$07$-$08 to 2016$-$08$-$09.
In their results they report the detection of multiple absorptions features in the X-ray spectra. They interpret one of these absorption features, detected at an observed-frame energy of $E$ = 8.6~keV, as originating from absorption by highly ionized iron outflowing with a speed of 0.236$c$ $\pm$ 0.006$c$. They report that the velocity of this wind does not vary with luminosity and time and the equivalent width (EW) of the absorption feature is anti-correlated to the 3$-$10~keV flux (see Figure 3 of \cite{2017MNRAS.469.1553P}).
The $v$ vs. $L_{\rm X}$ and $EW$ vs. $F_{\rm X}$ reported behavior of \iras\ appears to be very different from that detected in other luminous AGN such as $z=3.91$ APM 08279$+$5255 \citep{2002ApJ...579..169C,2009ApJ...706..644C,2011ApJ...737...91S}, $z=0.184$ PDS~456 \citep{2015Sci...347..860N,2017MNRAS.472L..15M, 2018ApJ...854L...8R}, $z=0.062$ PG~1126$-$041 \citep{2011A&A...536A..49G} and $z=2.7348$ HS~1700+6416 \citep{2012A&A...544A...2L}.
The relativistic outflow velocities detected in these AGN have been reported to vary with time and in some instances with luminosity.
A correlation has been found between the outflow velocity and the X-ray luminosity in APM 08279$+$5255 and PDS~456 (see Figure 10 in \cite{2011ApJ...737...91S} and Figure 3 in \cite{2017MNRAS.472L..15M}) suggesting that radiative driving may be contributing to the acceleration of these winds.
 \cite{2017Natur.543...83P} independently analyzed the variability of  \iras\ using principal component analysis (PCA).
 They report the detection of significant spectral variability at energies similar to the ones found in their previous spectral analysis.

In Section~2  we present the X-ray observations and data reduction of \iras. 
In Section~3 we  apply PCA to \iras\ following the methodology described in  \cite{2017Natur.543...83P}.
In Section~4 we search for possible variability of the outflow velocity as a function of flux and time.
In Section~5 we present our results on the dependence of the equivalent width of the iron absorption line with the 3$-$10~keV flux.
Finally, in Section~6 we present a summary of our conclusions.
Throughout this paper we adopt a flat $\Lambda$ cosmology with 
$H_{0}$ = 68~km~s$^{-1}$~Mpc$^{-1}$ $\Omega_{\rm \Lambda}$ = 0.69, and  $\Omega_{\rm M}$ = 0.31 \citep{2016A&A...594A..13P}.

\section{X-ray Observations and Data Reduction} \label{sec:data}

In Table \ref{tab:obslog} we list the observation dates, exposure times, the background-subtracted source count-rates and the 3$-$10~keV fluxes of \iras.
For the present analysis we are mostly concerned with examining the properties of absorption features with energies above 0.5~keV.
In this energy range the EPIC-pn has superior effective area than the two MOS detectors combined.
We therefore concentrated our effort in the reduction and analysis of the EPIC-pn data of \iras\ alone.

For the reduction of the \xmm\ observations we filtered the EPIC-pn \citep{2001A&A...365L..18S} data by selecting events corresponding to instrument \verb+PATTERNS+ in the 0--4 range (single and double pixel events). 
Several moderate-amplitude background flares were present during several of the \xmm\ observations. The EPIC-pn data were filtered on a rate of $< 20~{\rm cnts~s}^{-1}$,
using the SAS task \verb+tabgtigen+,  to exclude times when these flares occurred resulting in the effective exposure times listed in Table \ref{tab:obslog}. To test for sensitivity to background non-uniformity we also tried different background extraction regions. We did not find any differences in the spectral shapes and features using  more conservative threshold cuts or selecting different background extraction regions. The Epic-pn spectra were binned to have at least 20 counts per bin, as appropriate for fitting spectra using $\chi^{2}$ statistics.

The EPIC-pn observations of \iras\ were performed in Large Window mode to reduce the effects of pile-up.
The``tolerant" EPIC-pn flux limit threshold for Large Window mode is 6 counts~s$^{-1}$ \citep{2015A&A...581A.104J}.
According to this study, the fractional flux loss in Large Window mode and for soft spectra with maximum photon rates of 6 cnts~s$^{-1}$ is $\sim$ 6\%.
For the purpose of estimating the effects of pile-up we calculated the count rates over the entire energy band of the Epic-pn for circular source extraction regions with radii of 700 pu.
The maximum EPIC-pn count rates of spectra stacked by flux and by time in our analysis are 5.7 count s$^{-1}$ and 4.08 count s$^{-1}$, respectively.
Spectra stacked by flux and time have count rates below the ``tolerant" EPIC-pn flux limit threshold for Large Window mode and therefore flux loss  and spectral distortion due to
pile-up effects are negligible in our analysis. Moreover, \iras\ has a soft spectrum and the effects of pile-up are expected to be reduced at energies above $\sim$2~keV.
\cite{2017MNRAS.469.1553P} also investigated the effects of pile-up for these Epic-pn observations of \iras\ and did not find any spectral distortion
above 2~keV in this soft X-ray source.
We conclude that pile-up does not affect the results of our analysis, especially in the 3$-$10~keV band where we have detected absorption lines from the ultrafast outflow.

\section{Principal Component Analysis of \iras} \label{sec:pca}

We re-analyzed the 2016 {\sl XMM-Newton} observations of \iras\ using PCA as described in \cite{2017Natur.543...83P} to check if we could reproduce their published results. We tested the sensitivity of the reduction of the \iras\ data to the size and location of the background and source extraction regions. We also investigated the level of background at energies near the reported absorption features in both individual (used in the PCA analysis) and stacked spectra.

We select circular source extraction regions of two different radii, $r$ = 250 and 600 pn physical units  (pu, 1~pu = 0\sarc05) centered on \iras\  and circular background regions of $r$ = 1600~pu,  making sure that the background extraction regions were outside of the elevated Cu background ring region of the EPIC-pn.
The source extraction region of $r$ = 600~pu was selected to compare with the results published in \cite{2017MNRAS.469.1553P}.
Specifically,  \cite{2017MNRAS.469.1553P} show their source and background regions in their Extended Data Figure 2 marked by small and large white circles, respectively. From their Extended Data Figure 2 and the known angular size of the Epic-pn CCD's we infer that the source and background diameters of their extractions regions are 60\arcsec\ and 120\arcsec, respectively.  
We also determined the 0.3$-$10~keV light-curves of \iras\ for extraction regions of diameter  60\arcsec\ to compare with the 0.3$-$10 keV light-curve presented in their Extended Figure 1 of \cite{2017MNRAS.469.1553P}. We  reproduce the same count rates as in their Extended Figure 1 when using 60\arcsec\ diameter circular extraction source extraction regions.

We selected a source extraction region of $r$ = 250~pu to optimize the S/N ratio in the 8$-$10~keV region which contains the detected blueshifted \FeXXV\ and  \FeXXVI\ resonance absorption lines. 
The percent decrease in the background subtracted source counts in the 0.4$-$10~keV band obtained by using the  $r$~=~250~pu extraction region compared to the 600~pu region is about 20\%. However, the background of spectra extracted from the 250~pu regions is reduced by a factor of $\sim$ 6 compared to spectra extracted using the 600~pu regions.
A more detailed justification of the selection of the $r$ = 250~pu source extraction regions is provided later on in section \ref{sec:variability}.

We employed the {\sl XMM-Newton} science analysis system tool \verb+arfgen+  with enabled encircled energy correction for point sources to create the ancillary response files (arf) appropriate for each selected source extraction region. 

Observations were divided into increments of $t$ = 10~ks to match the intervals used in the \cite{2017Natur.543...83P} analysis
and pn source and background spectra were extracted for each 10~ks interval.  
For the pn data reduction we used SAS software version 16.1.

The background subtracted spectra were rebined into equal logarithmic bins in energy and the PCA analysis was performed
for a range of fractional bin-widths. 
The 10~ks spectra were stored in a matrix $f(t_{\rm j}, E_{\rm i})$, where  the $n$ columns are the energies $E_{\rm i}$ of the spectral bins and the $m$ rows are the times $t_{\rm j}$ of the 10~ks intervals. 

The average spectrum over all $m$ 10~ks  intervals was computed as, 

\begin{equation}
f_{ ave}{\left( { E }_{ i } \right)}=\frac { \sum _{ j=1 }^{ m }{ f\left( { t }_{ j },{ E }_{ i } \right)  }  }{ m } 
\end{equation}

The normalized variation of a spectrum from the average spectrum $f_{\rm ave}$ was calculated as 

\begin{equation}
f_{\rm var}\left( { t }_{ j },{ E }_{ i } \right) =\frac { f\left( { t }_{ j },{ E }_{ i } \right) -f_{ ave}{\left( { E }_{ i } \right)} }{ f_{ ave}{\left( { E }_{ i } \right)} } 
\end{equation}

Using the IDL routine SVDC we computed the singular value decomposition of  the m $\times$ n  matrix $f_{\rm var}(t_{\rm j},E_{\rm i})$
as the product of an (m $\times$ n) orthogonal array A, an (n $\times$ n) diagonal array SV, composed of the singular values, and the transpose of an (n $\times$ n) orthogonal array PC.

\begin{figure}
\plotone{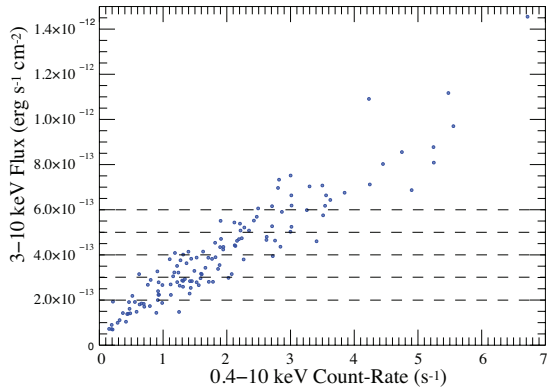}
\caption{The 3$-$10~keV fluxes of the 10~ks intervals of  \iras\ as a function of the 0.4$-$10~keV count-rates.  The horizontal dashed lines indicate the boundaries of the flux ranges selected for stacking spectra.\label{fig:fluxranges}}
\end{figure}

\begin{figure}
\plotone{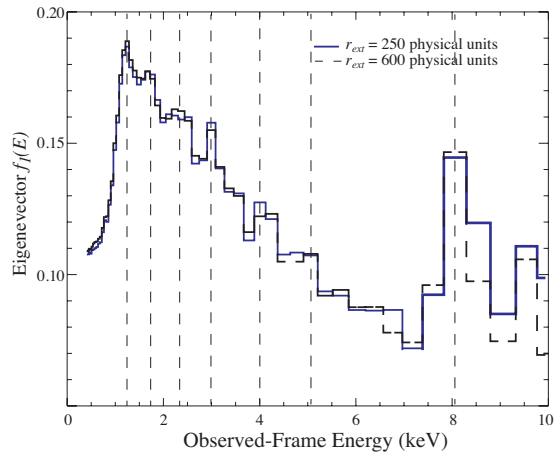}
\caption{The first eigenvector component $PC(1,E)$ from PCA applied to  \iras.  Spectra are extracted from circular source regions of $r_{\rm ext}$ = 600 physical units (30\arcsec) and $r_{\rm ext}$ = 250 physical units (12\sarc5). The vertical lines indicate the energies of the  peaks in the variability.  \label{fig:pca}}
\end{figure}

\begin{figure*}[ht!]
\plotone{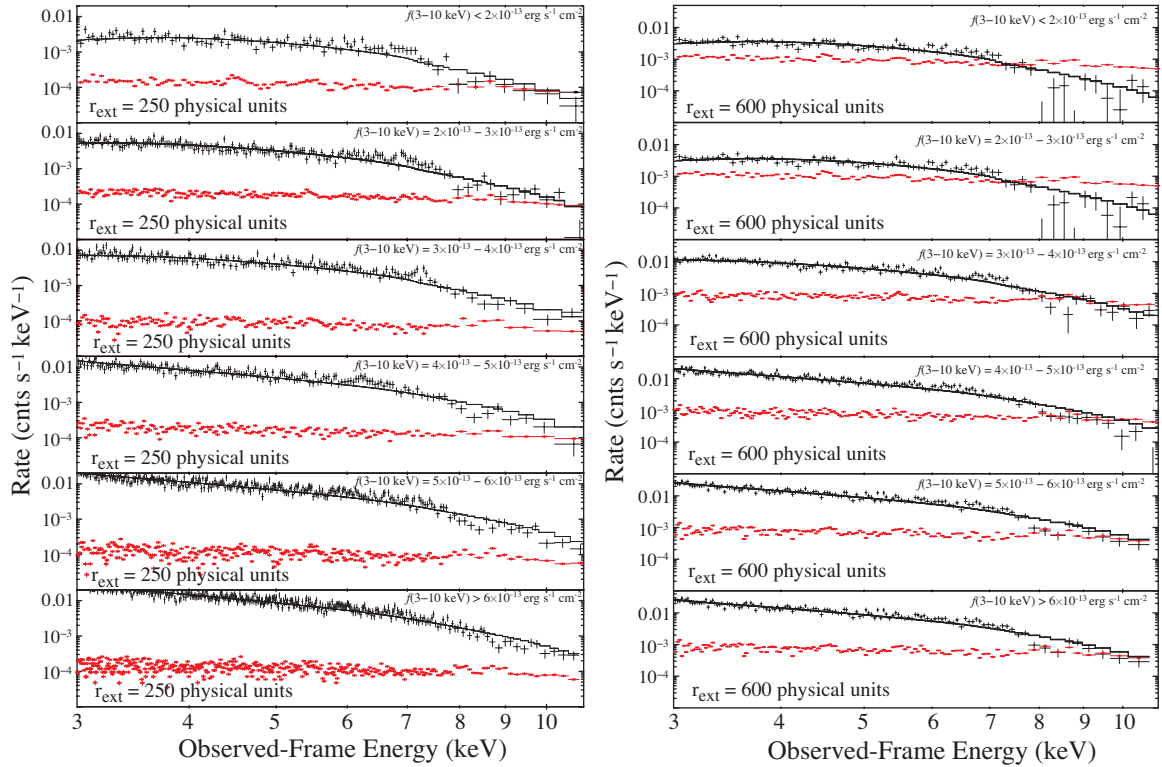}
\caption{ The EPIC-pn spectra of \iras\  combined over several ranges of 3$-$10~keV fluxes listed in the panels. The spectra are fit with a simple absorbed power-law model to illustrate the presence of possible emission and absorption features. The red data points represent the background spectra. The black data points represent the net source spectra (source minus background).  Left: Spectra are extracted from circular source regions of  $r_{\rm ext}$ = 250 physical units (12\sarc5). Right: Spectra are extracted from circular source regions of $r_{\rm ext}$ = 600 physical units (30\arcsec).   \label{fig:bkg}}
\end{figure*}

\begin{equation}
f_{\rm var}=(A)(SV)({PC) }^{ T }
\end{equation}

The spectral variability as a function of energy is stored in the principal components  PC($i$,$E _{ i }$), where most of the variability of the source 
is stored in the first component PC(1,$E _{ i }$). The contributions of principal components to each 10~ks spectrum are stored in the light-curves  A($i$,$t _{j}$).

\begin{equation}
f_{\rm var}\left( { t }_{ j },{ E }_{ i } \right) =\sum _{ i=1 }^{ n }{ A(i,{ t }_{ j }) } PC(i,{ E }_{ i })
\end{equation}

\cite{2017MNRAS.469.1553P} stacked spectra that fell within a defined range of 0.3$-$10~keV count-rates and determined the 3$-$10~keV flux of each stacked spectrum.  
We initially followed this approach by stacking spectra as a function of count rates (see Table \ref{tab:counts}) to compare results but we also adopted a slightly different approach in selecting how to stack spectra.  Specifically, we determined the individual 3$-$10~keV fluxes of each 10~ks spectrum, stacked spectra that fell within a defined range of 3$-$10~keV fluxes and determined the 3$-$10~keV flux of each stacked spectrum. We chose this approach since count-rates and 3$-$10~keV fluxes are not exactly proportional in the presence of spectral variability. This is shown in Figure \ref{fig:fluxranges} where we show a plot of total count-rates versus 3$-$10~keV fluxes for all 10~ks spectra of \iras.  Considerable scatter between  3$-$10~keV fluxes and 0.3$-$10~keV count-rates is evident. We conclude that the inferred 3$-$10~keV fluxes of stacked spectra that were combined based on their total count-rates do not accurately represent the fluxes of the stacked spectra.

We next investigated the sensitivity of the reported PCA results on \iras\ to background. The first eigenvector component $PC(1,E)$ from PCA obtained from data extracted from circular source extraction regions of $r_{\rm ext}$ = 600~pu (30\arcsec) and $r_{\rm ext}$~=~250~pu (12\sarc5) radii are presented in Figure \ref{fig:pca}. We confirm the enhanced spectral variability at energies similar to the ones reported in \cite{2017Natur.543...83P}. The only difference is a slight increase of $PC(1,E)$  for energies above observed-frame  energies of 8~keV, when background contamination is reduced.

\section{VARIABILITY OF THE OUTFLOW OF \iras.} \label{sec:variability}

\subsection{Variability of Outflow as a Function of  3$-$10~keV Flux}

In Figure \ref{fig:bkg} we plot the combined EPIC-pn spectra of \iras\ extracted from circular source extraction regions of $r_{\rm ext}$ = 600~pu (referred to as $r_{600}$ spectra) and over-plot the background spectra. The spectra were combined in 
six flux regimes based on their 0.3$-$10~keV fluxes listed in the panels of Figure \ref{fig:bkg}.

We find that the background begins to dominate at observed-frame energies above $\sim$ 7.5~keV. In Figure \ref{fig:bkg} we also plot the same spectra of \iras, however, extracted from circular source regions of $r_{\rm ext}$ = 250~pu (referred to as $r_{250}$ spectra). The background in all flux levels for the $r_{250}$ spectra is significantly reduced compared to the $r_{600}$ spectra.

\begin{figure}
\plotone{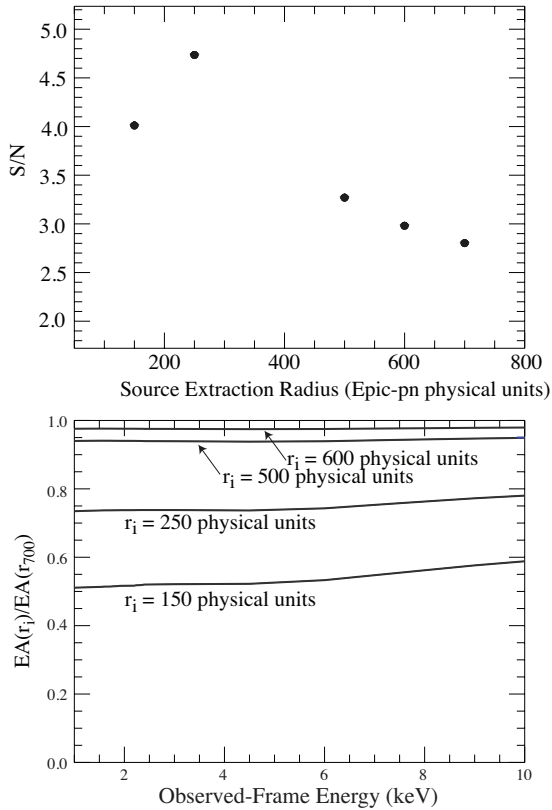}
\caption{ Top: The S/N ratio of EPIC-pn spectra of \iras\  calculated in the 8$-$10~keV observed-frame energy band as a function of source extraction radius. The comparison was made between combined pn spectra in the lowest flux regime shown in Figure \ref{fig:bkg}.   Bottom: The ratio of the pn effective area using source extraction regions of 
$r$~=~150, 250, 500, and 600 pn physical units to that of a $r$~=~700 physical units source extraction region (1 physical unit = 0\sarc05.  \label{fig:arfrat}}
\end{figure}

To justify the selection of an extraction region of  $r_{\rm ext}$~=~250~pu we show in Figure \ref{fig:arfrat} the S/N ratios in the 8$-$10~keV region of \iras\ for source extraction regions of 150, 250, 500, 600, and 700~pu and background subtraction regions of $r$ = 1600~pu. The S/N ratios were calculated for combined spectra of \iras\ with 3$-$10~keV fluxes of less than 2 $\times$ 10$^{-13}$~erg~s$^{-1}$~cm$^{-2}$, the lowest  3$-$10~keV flux range shown in Figure \ref{fig:bkg}.
A significant improvement in the S/N ratios of the spectra in the 8$-$10~keV range is achieved by selecting extraction regions of  $r_{\rm ext}$ = 250~pu
compared to  $r_{\rm ext}$ = 600~pu. In Figure~\ref{fig:arfrat} we also plot the ratios of the pn effective area using $r$~=~150, 250, 500, and 600~pu source extraction regions to that of a $r$ = 700~pu source extraction region. We note that the effective-area ratios are smooth functions across the 8$-$10~keV observed-frame energy range.

\begin{figure}[ht!]
\plotone{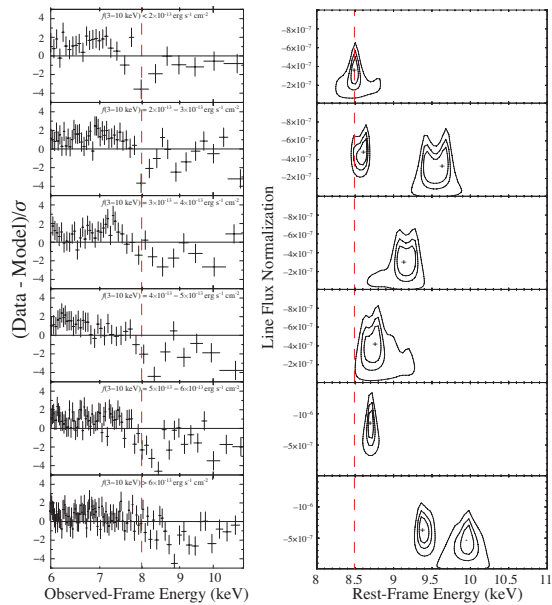}
\caption{Left: Residuals of fits of a simple absorbed power-law to the $r_{250}$ spectra (stacked by 3$-$10keV flux) of \iras\ shown in Figure \ref{fig:bkg}. Significant absorption lines are detected between the observed-frame energies of  8 and 10~keV. Right: 68\%, 90\%, and 99\% confidence contours of the rest-frame energies of the detected absorption lines versus their line strengths. To produce the confidence contours we fit the stacked $r_{250}$ spectra with a simple absorbed power-law model that contains one or two absorption lines and up to one emission line.  For comparison purposes we show vertical lines at the energy of the best-fit absorption line corresponding to the lowest flux state.   \label{fig:res}}
\end{figure}

We conclude that modeling the EPIC-pn spectra of \iras\ extracted from circular source extraction regions of $r_{\rm ext}$ = 600~pu will likely result in unreliable results especially for features detected above observed-frame energies of $\sim$ 7~keV.
To illustrate the presence of possible emission and/or absorption features we fit the stacked $r_{250}$ 
spectra of \iras\ with a simple absorbed power-law model. Clear residual features are detected in the stacked $r_{250}$ spectra of \iras\ (see Figures \ref{fig:bkg} and \ref{fig:res}).

Guided by these residual features we next fit the stacked $r_{250}$ spectra with a simple absorbed power-law model that contains
one or two absorption lines and up to two emission lines. Since we are only interested in constraining the high-energy absorption features we restricted the spectral fits to the energy range of 3--11~keV. The 68\%, 90\%, and 99\% confidence contours of the rest-frame energies of the detected absorption lines versus their line strengths are shown in Figure \ref{fig:res}. Only absorption lines detected at the  $>$ 99\%  confidence level are presented in this figure. Significant high-energy absorption features are detected in all flux ranges. The energies of the absorption lines vary with 3$-$10~keV flux ranging between rest-frame energies of  8.64$_{-0.06}^{+0.05}$~keV and   9.96$_{-0.10}^{+0.09}$~keV.  As we show later on in this section these absorption lines are likely the result of absorption by highly ionized iron (\FeXXV\ and \FeXXVI).


We followed a more robust approach of estimating the significance of the blueshifted absorption iron lines based on Monte Carlo simulations to determine the distribution of the $F$-statistic between different models \citep{2002ApJ...571..545P}.  Following this approach, for each spectrum we simulated 1000 data sets using the XSPEC \verb+fakeit+ command \citep{1996ASPC..101...17A}. 
We considered a null model that included a simple absorbed power-law and an alternative model that in addition included one or two Gaussian absorption lines.
We fit the 1000 data sets with the null and alternative models and determined the distribution of the 
$F$-statistic\footnote{The $F$-statistic is given by $F=\frac { { \chi  }_{ { \nu  }_{ 1 } }^{ 2 }-{ \chi  }_{ { \nu  }_{ 2 } }^{ 2 } }{ \Delta \nu  } /\frac { { \chi  }_{ { \nu  }_{ 2 } }^{ 2 } }{ { \nu  }_{ 2 } } $}  
from these fits. Finally, we computed the probability for the $F$ value to exceed the value determined from the fits of the null and alternative models to the observed spectra. The Monte Carlo simulations 
indicate that the shifted iron lines are detected at  $>$ 99\% confidence in all flux levels and confirm the significance of the detections calculated using $\chi^{2}$ confidence contours. In Table \ref{tab:linepro_f} we present the energies and the significance of the iron absorption lines detected in \iras\ as a function of the 3$-$10~keV flux of the $r_{250}$ spectra that were combined in selected ranges of fluxes shown in Figure \ref{fig:fluxranges}.

\subsection{Variability of Outflow as a Function of  Time.}
\begin{figure*}[ht!]
\plotone{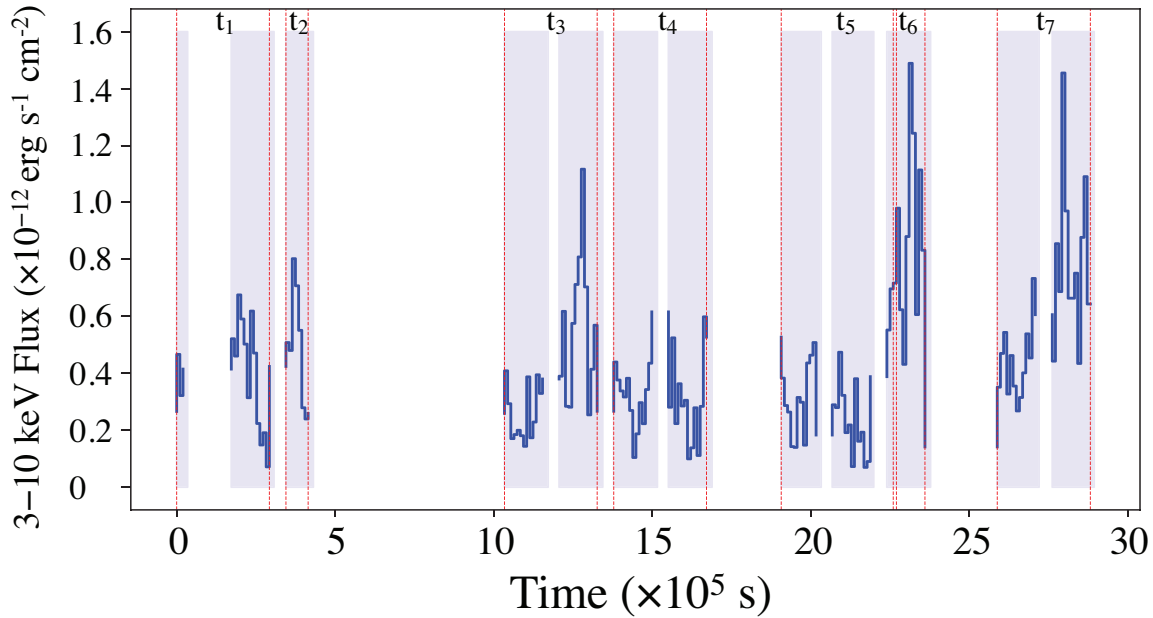}
\caption{The light-curve of the 3$-$10~keV flux of \iras\ in increments of 10~ks time bins. The vertical dashed lines represent the boundaries of the 7 time intervals within which the 10~ks $r_{250}$ spectra were stacked. The shaded regions represent the twelve observations listed in Table \ref{tab:obslog}.  \label{fig:lc}}
\end{figure*}

We also investigated the variability of the outflow as a function of time. In Figure \ref{fig:lc} we show the light-curve of the 3$-$10~keV flux of \iras. We selected 7 consecutive time intervals to study the evolution of the outflow and combined the EPIC-pn $r_{250}$ spectra in these intervals. The selection of the time intervals in our time resolved spectroscopic analysis was based on the following criteria. We chose consecutive time intervals with similar S/N spectra to allow for spectral fitting using photoionization codes. We also selected time intervals that isolated large flares.
Specifically, in several of these intervals significant flares are evident where the 3$-$10~keV flux varies by a factor of $\sim$2 in a timescale as short as 10~ks which is consistent with a light crossing-time size of less than $\sim$ 300~$r_{\rm g}$, where $r_{\rm g}$~=~$GM_{\rm BH}/c^{2}$
and $M_{\rm BH}$ $\sim$ 6 $\times$ 10$^{6}$ M$_{\odot}$  is the black hole mass of \iras\ reported in \citep{2005ApJ...618L..83Z}. 
The published values of the black hole mass of \iras\ show modest differences based on the method used to estimate them. For example, by measuring the accretion disk flux, \cite{2015MNRAS.446..759C} obtained $M_{\rm BH}$~=~3.5$_{-0.6}^{+5.5}$~$\times$~10$^{6}$ M$_{\odot}$,
whereas reverberation analysis presented in \cite{2014MNRAS.439.3931E} reports $M_{\rm BH}$~=~9.3$_{-2.9}^{+3.4}$~$\times$~10$^{6}$ M$_{\odot}$.

\begin{figure}[ht!]
\plotone{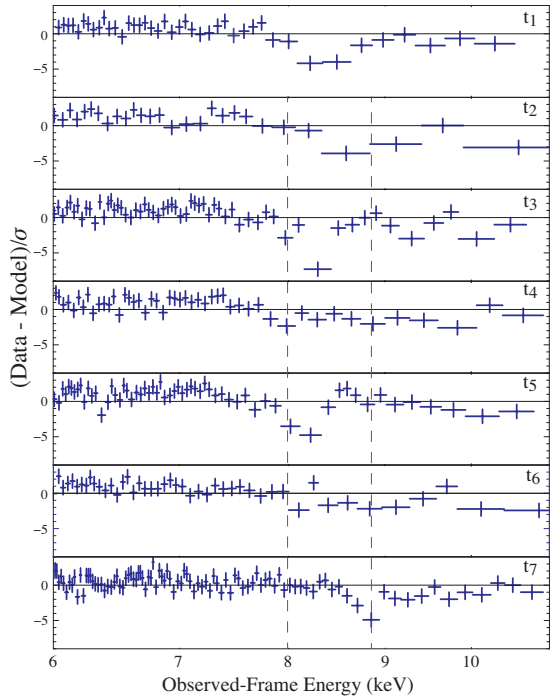}
\caption{The residuals of fits of a simple absorbed power-law to the stacked $r_{250}$ spectra of \iras\  over the time intervals shown in Figure \ref{fig:lc}. The time evolution of the velocity of the relativistic outflow is clearly evident. The vertical dashed lines indicate the energies of the minimum and maximum blueshifted absorption lines. \label{fig:res_time}} 
\end{figure}

In Figure \ref{fig:res_time} we show the residuals between a fitted model and the spectra in the 6$-$11~keV observed-energy range. The model used to produce these residuals is a simple absorbed power-law model. We verified that the background spectra lie below all the stacked $r_{250}$ source spectra for energies below $\sim$ 10~keV. The reduced background of the $r_{250}$ spectra allows for reliable identifications of the high-energy absorption lines. We find significant variability of the energy and velocity of the outflow of \iras. 

\begin{figure}[ht!]
\plotone{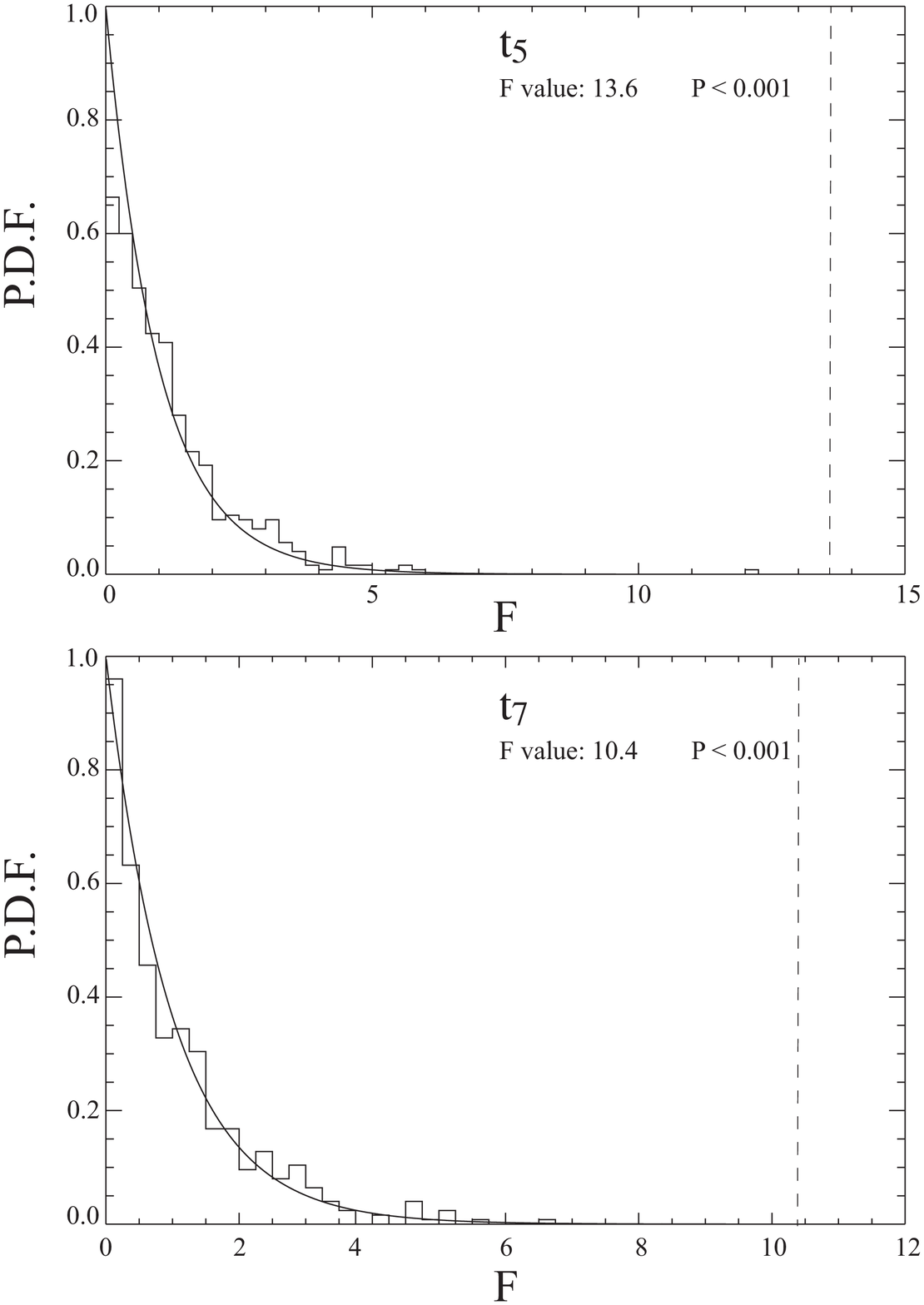}
\caption{Monte Carlo simulated (histogram) and theoretical (smooth curve) probability density distributions of the $F$-statistic between fits of models that included a simple absorbed power-law (null model), and one that included one Gaussian absorption line (alternative model) to the observed spectra of \iras\ during intervals $t_{5}$ (top) and $t_{7}$ (bottom). The results of the Monte Carlo simulations for all 7 time intervals are presented in Table \ref{tab:linepro_t}. \label{fig:f_test}} 
\end{figure}

In order to determine the energies and significance of the absorption lines in the seven time intervals we fit the $r_{250}$ spectra of \iras\ with a model that consists
of a simple absorbed power-law, a Gaussian absorption line and one or two emission lines. 
In Table \ref{tab:linepro_t} we present the energies and the significance of the iron absorption lines detected in \iras\ for the seven time intervals
shown in Figure \ref{fig:lc}. The most significant change of the outflow velocity of \iras\  is observed between intervals $t_{5}$ and $t_{7}$.
In Figure \ref{fig:f_test} we show a typical example of the Monte Carlo simulated distribution of the $F$-statistic between fits of the null and alternative models to the observed spectra of \iras\ obtained during intervals  $t_{5}$ and $t_{7}$. The blueshifted lines in these time intervals  are detected at the $>$ 99.9\% confidence.

We also searched for changes of the energies of possible soft X-ray absorption lines by fitting the full band 0.5$-$10~keV Epic-pn spectra of the time intervals $t_{5}$ and $t_{7}$ that showed the most dramatic variability of the energies of the Fe absorption lines. The model fit to the full energy band was similar to the one used in the \cite{2018MNRAS.476.1021P} analysis. Specifically, we used a model that consists of Galactic absorption, intrinsic cold absorption, a blackbody to account for the soft excess, a power-law for the direct emission, a relativistically blurred reflected emission component and X-ray transmission through a photo-ionized gas.
For the relativistically blurred X-ray reflection we used the \verb+RELXILL+ model \citep{2014MNRAS.444L.100D,2014ApJ...782...76G}. 
We use the analytic \verb+XSTAR+ model \verb+warmabs+ \citep{2001ApJS..133..221K} to model the transmission of the outflowing intrinsic ionized absorber.

The fit with this model shows no significant features between 1.5$-$10~keV except for the $\FeXXV/\FeXXVI$ lines, however, several broad residual features are present between 0.5$-$1.5~keV.
The origin of these broad residual features is uncertain, however, these features may result from modeling the soft excess with a simple blackbody.
Other plausible candidates for these broad residuals include pile-up that is known to distort spectra and will be more significant at energies below $\sim$2~keV, and the presence of
multiple outflowing components of various ionization states.

\subsection{Correlation Between Outflow Velocity and 3$-$10~keV Flux.}
\begin{figure}[ht!]
\plotone{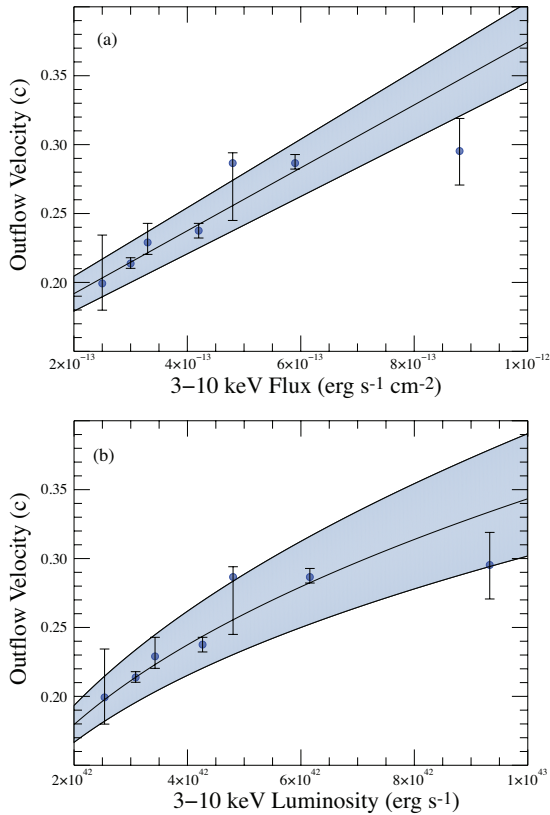}
\caption{ The outflow velocity of the ionized absorber as a function of the 3$-$10~keV flux (a) and 3$-$10~keV Luminosity (b) of \iras. The outflow velocities were determined by fitting Gaussians to the absorption lines detected in the $r_{250}$ spectra of the seven time intervals shown in Figure \ref{fig:lc}. We also show the straight-line least-squares fit (a) and the power-law least-squares fit (b) to the data in the solid lines. The shaded area represents the uncertainty of our fits to the $r_{250}$ data. \label{fig:v_vs_flux}}
\end{figure}

We searched for a possible dependence between the outflow velocity and the 3$-$10~keV flux in the seven time intervals shown in Figure \ref{fig:lc}.  
The line energies listed in Table \ref{tab:linepro_f} and obtained from fitting the $r_{250}$ spectra with a model that consisted of a simple absorbed power-law and one or two Gaussian absorption lines
were converted to outflow velocities. Specifically, for the conversion we used the relativistic Doppler formula assuming that the high-energy absorption lines are the result of resonance absorption from ions of \FeXXVI\ in a gas with solar abundances, and the angle between our line of sight and the outflow direction is zero.
In Figure \ref{fig:v_vs_flux} we show that a trend exists, with the outflow velocity increasing with 3$-$10~keV flux. 
We fit a null hypothesis model of a constant behavior between velocity vs. flux and find $\chi^2 = 113$ for 6 degrees of freedom ($\nu$).
The probability $P_{\chi}$($\chi^2$;$\nu$) of exceeding $\chi^2 = 113$ for $\nu$ = 6 is $P_{\chi}(113;6) < 1 \times 10^{-15}$.
We conclude that the null hypothesis model of a constant behavior between velocity vs. flux is rejected at a very high significance level.
Fits of straight-line and power-law models to the velocity vs. flux data resulted in $\chi^2/\nu$ = 8.6/5 and $\chi^2/\nu$ = 6.8/5, respectively.
Similar trends have been reported in APM~08279$+$5255 and PDS~456 \citep{2011ApJ...737...91S, 2017MNRAS.472L..15M}.  We find a correlation between $v$ and the 3$-$10~keV flux with a Kendall's rank correlation coefficient of $\tau$ =  0.98 significant at $>$~99.8\% confidence. 
We use the following equation to describe the basic dynamics of a radiation-driven outflow \citep[e.g., equation 1 of][]{2002ApJ...579..169C}:

\begin{equation} \label{eq:vwi}
v_{\rm wind}=\left[\right( \Gamma_{\rm f} \frac{L_{\rm Bol}}{L_{\rm Edd}}-1
\left) \right ( \frac{1}{R_{\rm launch}}-\frac{1}{R} \left ) \right]^{1/2}
\end{equation}
where $v_{\rm wind}$ is the outflow velocity in units of $c$,
$\Gamma_f$ is the force multiplier, $L_{\rm Bol}$ and $L_{\rm Edd}$ are the Bolometric and Eddington
luminosities, respectively, $R_{\rm launch}$ is the radius (units of $R_S$) at
which the wind is launched from the disk, and $R$ is the distance
(units of $R_S$) from the central source. 

We fit the outflow velocity versus 3$-$10~keV luminosity of \iras\ with the function $v_{wind}$ =  $aL^{b}$ to compare the observed trend with Equation \ref{eq:vwi}. 
We find a best-fit value of $b = 0.40 \pm 0.03$ with a $\chi^2/\nu$ = 1.3. A fit with a straight line results in a significantly worse fit with  $\chi^2/\nu = 1.7$.
The best-fit value of $b = 0.40 \pm 0.03$ would imply that radiation driving may be contributing to the acceleration of this wind, however, there are several concerns	with this interpretation that we	
describe later on in this section. We note that a similar trend was found in PDS~456 with a best-fit value of $b = 0.20 \pm 0.05$.
The best-fit model $v_{wind}$ =  $aL^{b}$ fit to the outflow velocity versus 3$-$10~keV luminosity of \iras\ is shown in Figure  \ref{fig:v_vs_flux}.

\cite{2018MNRAS.476.1021P} reported a tentative correlation between the outflow velocity and the luminosity of \iras\ based on modeling spectra stacked by count-rate levels 
and extracted from circular regions with $r$~=~600~pu. The highly ionized iron line, however, is not detected in their analysis to shift with luminosity and they find the strength of the iron absorption line to decrease with luminosity.  \cite{2018MNRAS.476.1021P} conclude that the ultrafast outflow disappears in high-radiations fields.  
We have shown that the $r_{600}$ spectra of \iras\ are background dominated above $\sim$7~keV (see Figure  \ref{fig:bkg}) and that the analysis of the $r_{250}$ spectra of \iras,
that are not background dominated, clearly show the increase in the energies of the \FeXXV\ and \FeXXV\ lines with increasing 3$-$10~keV flux. 
As we show in section \ref{sec:width}, the equivalent width of the Fe line does not decrease with luminosity when analyzing the $r_{250}$ spectra of \iras\ and the 
ultrafast outflow, as manifested in the blueshifted highly ionized iron lines, does not vanish in the high radiation fields.
One possible reason these shifts of the Fe line were not detected previously is that in the high flux states the Fe lines shift into even more background dominated parts of the 
$r_{600}$ spectra of \iras. 

It has been suggested that similar correlations found in  APM~08279$+$5255 and PDS~456 may indicate that radiation pressure is contributing to the acceleration of the wind since the terminal velocity of a radiatively driven wind scales as  ${ v }_{ \infty  }\propto \sqrt { { \Gamma  }_{ f }L } $, where ${ \Gamma  }_{ f }$ is the force multiplier of the outflowing absorber.  
However, there are some problems with the premise that the main acceleration mechanism for these relativistic winds is radiation driving. Specifically, 
the launching radii inferred from the large relativistic outflow velocities of $\simlt$ 100~$r_{\rm g}$, indicate that only a fraction of the UV emitting region of the accretion disk will be contributing to radiation driving at these radii. Moreover, the terminal velocity of the outflowing wind depends on the force multiplier which decreases with increasing ionization parameter. As shown in Figure A.6 of \cite{2011ApJ...737...91S} for the best-fit values of the ionization parameters found in \iras\ the force multiplier is calculated to be close to 1, implying that radiation driving, at least near the launching site, is not a main contributor to the acceleration of the wind. Finally, the 3$-$10~keV luminosity of AGN is a relatively small fraction of the bolometric luminosity and the central X-ray flux alone cannot accelerate the wind to the observed speeds and result in the calculated  efficiencies that we find for \iras\  and are presented in the end of this section.

We next proceed with fitting the spectra of the intervals $t_{5}$ and $t_{7}$ to determine the properties of the outflowing absorber and search for possible 
changes of the absorber properties over a period of $\sim$~500~ks. Guided by the shape and location of identified absorption and emission residuals (see Figure \ref{fig:res_time}) 
we fit the spectra of intervals $t_{5}$ and $t_{7}$ with a model that  consists of a power-law modified by Galactic absorption, an outflowing intrinsic ionized absorber and
relativistically blurred X-ray reflection from parts of the accretion disk near the black hole.
We attempt to mimic the velocity broadening of the absorption lines by introducing in the \verb+XSTAR+ \verb+warmabs+ model turbulent velocities.
We performed several fits using  \verb+warmabs+ where we allowed the turbulence velocity to vary and found best-fit values of
$v_{\rm turb}$ $\sim$ 5,500~km~s$^{-1}$ for interval $t_{5}$ and $v_{\rm turb}$ $\sim$ 1,500~km~s$^{-1}$ for interval $t_{7}$. 
The spectral fits were performed in the 2 $-$ 10~keV observed-frame energy range to avoid the soft excess at lower energies and to focus on the blueshifted absorption lines in this range. The default atomic population file \verb+pops.fits+  provided in NASA's \verb+warmabs+ distribution uses a fixed value of the photon index of  $\Gamma$~=~2. However, our spectral analysis indicates that the 2$-$10~keV photon index $\Gamma$ increased from $\sim$ 2.0 to $\sim$ 2.4 between intervals $t_{5}$ and $t_{7}$. We therefore used \verb+XSTAR+ to create new population files appropriate for photon indices  of the spectra of intervals $t_{5}$ and $t_{7}$. The best-fit values for the column density, ionization parameter and outflow velocity of the wind from fitting this model to the spectra of intervals $t_5$ and $t_7$ are presented in Table~\ref{tab:properties}.  We find a significant change in outflow velocity from $v$ = 0.173$_{-0.006}^{+0.005}$$c$  to $v$ = 0.283$_{-0.006}^{+0.006}$$c$  within a period of about 500~ks.
The velocities derived from the \verb+XSTAR+ photoionization model (listed in Table \ref{tab:properties}) are consistent, within errors, with the velocities derived from the spectral fits using Gaussian absorption lines and show a similar trend with flux. We note that that the outflow velocities derived from fits to the $r_{250}$ spectra using the \verb+relxill+ + \verb+warmabs+ model are not as sensitive (as compared to fitting Gaussian lines for example) to the assumed ionization state.

\begin{figure}[ht!]
\plotone{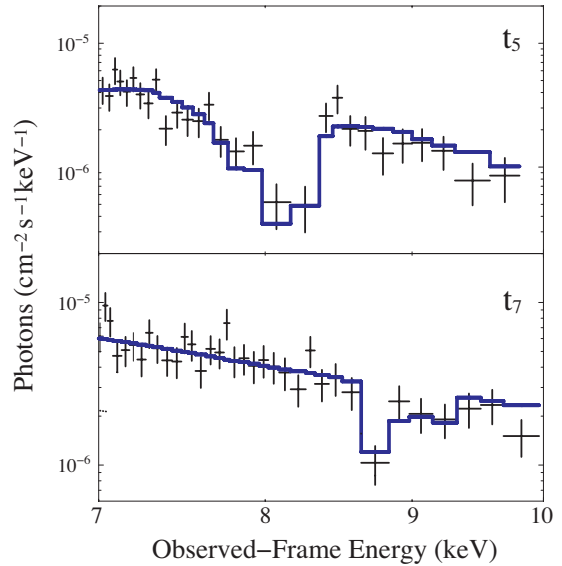}
\caption{The unfolded spectra and and best-fit photoionization models to the \iras\ $r_{250}$ spectra of intervals $t_{5}$ (top)  and $t_{7}$ (bottom). The inferred  outflow velocities of the absorber for intervals  $t_{5}$ and $t_{7}$  are 
$v$ = 0.173$_{-0.006}^{+0.005}$$c$  and $v$ = 0.283$_{-0.006}^{+0.006}$$c$, respectively.  \label{fig:unfolded}}
\end{figure}

\begin{figure}[ht!]
\plotone{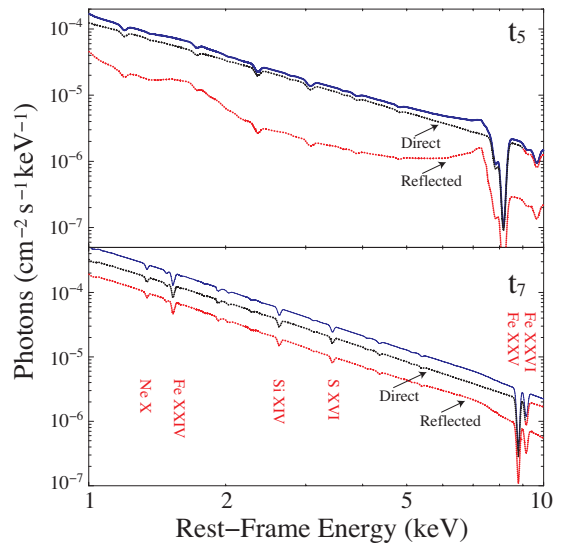}
\caption{The best-fit photoionization models to the \iras\ $r_{250}$ spectra of intervals $t_{5}$ (top)  and $t_{7}$ (bottom). 
The black and red dotted lines indicate the direct and reflected components, respectively.
The blueshifted resonance spectral lines of \NeX, \FeXXIV, \SiXIV, \SXVI, \FeXXV,  and \FeXXVI\ are indicated in red. \label{fig:models}}
\end{figure}

\begin{figure}[ht!]
\plotone{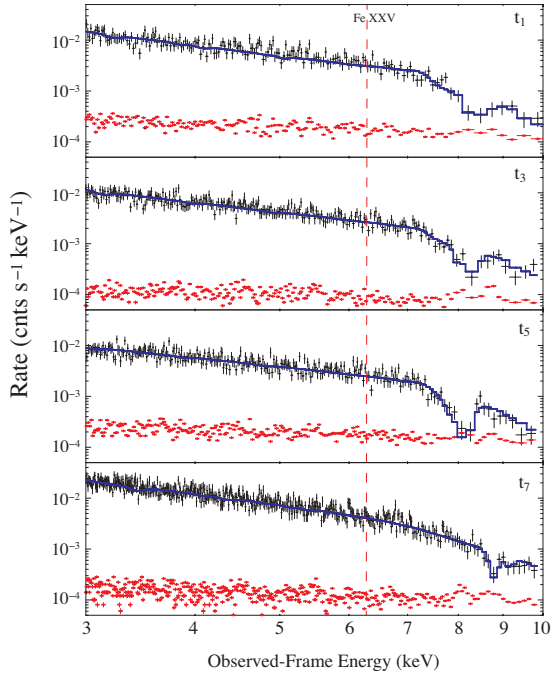}
\caption{ The $r_{250}$ spectra of \iras\ in the time intervals $t_{1}$, $t_{3}$, $t_{5}$, and $t_{7}$. The solid blue curve represents the best-fit model that  consists of a power-law modified by Galactic absorption, an outflowing intrinsic ionized absorber and relativistically blurred X-ray reflection. The red stars show the background spectra. The vertical dashed line shows the location of the rest-frame Fe XXV line. \label{fig:t1356}}
\end{figure}

In Figure \ref{fig:unfolded} we show the unfolded best-fit photoionization plus relativistic reflection models to the spectra of intervals $t_{5}$ and $t_{7}$ in the region near the blue-shifted \FeXXV\ and \FeXXVI\ absorption lines. The spectral evolution of the wind is evident between intervals $t_{5}$ and $t_{7}$. 
The best-fit photoionization models to the $r_{250}$ spectra of intervals $t_{5}$ and $t_{7}$ are shown in Figure \ref{fig:models}. Variability of the direct and reflected components
and the increase in outflow velocity are evident. Since Fe absorption lines were also detected at $>$~99\% confidence in time intervals $t_{1}$ and $t_{3}$ we fit the spectra of \iras\ in these intervals with the model that  consists of a power-law modified by Galactic absorption, an outflowing intrinsic ionized absorber and relativistically blurred X-ray reflection. 
The properties of the outflow for time intervals $t_{1}$ and $t_{3}$ are also included in Table~\ref{tab:properties}.  In Figure \ref{fig:t1356} we present the $r_{250}$ spectra of \iras\ in the time intervals $t_{1}$, $t_{3}$, $t_{5}$, and $t_{7}$, in which we have detected the Fe absorption line at $\simgt$ 99.9\% confidence. We over-plot the best-fit model that consists of a power-law modified by Galactic absorption, an outflowing intrinsic ionized absorber and relativistically blurred X-ray reflection. We also over-plot the background spectra to show that the detected Fe absorption lines are not affected by background. The $r_{250}$ stacked spectra clearly show the variability of strength and energy of the Fe line and show that the Fe line is detected at the high ionization levels as well (see Table 5).

\begin{figure}[ht!]
\plotone{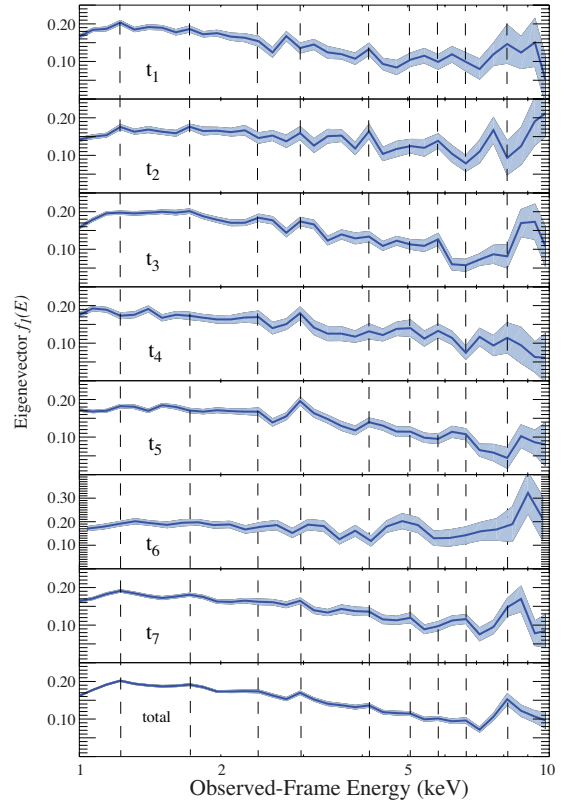}
\caption{Time-resolved PCA analysis of \iras. The first eigenvector component PC(1,E) from PCA applied to the 7 time intervals of the $r_{250}$ spectra of  \iras. The bottom panel shows the PCA applied to the total exposure time. The shaded regions represent the errors as calculated by performing a Monte Carlo analysis to the input $r_{250}$ spectra and taking the standard deviation of the results.  The vertical dashed lines indicate the locations of the peaks of PC(1,E) from the total exposure. \label{fig:pca6}}
\end{figure}

The observed variability of the energies of the absorption lines is likely to have contributed to the broadening of the peaks of the eigenvector component PC(1,E) shown in Figure \ref{fig:pca}. We searched for possible variability of component PC(1,E) by performing time-resolved PCA analysis to the seven time intervals shown in Figure \ref{fig:lc}. The inclusion of fewer spectra in the PCA analysis, when performed in each time interval, will lead to larger uncertainties due to Poisson noise. We therefore estimated the uncertainty of the PC(1,E) components in each time interval by performing a Monte Carlo analysis to the input spectra and taking the standard deviation of the results. A similar approach in estimating the errors in the PCA analysis was used by \cite{2008A&A...483..437M} and \cite{2017Natur.543...83P}.   
In Figure \ref{fig:pca6} we show the first eigenvector component PC(1,E) from PCA applied to the seven time intervals  of  \iras. 

The significance of several of the soft-X-ray PCA(1,E) peaks for the individual time intervals $t_{1}$ through $t_{7}$ is relatively low, and we focus our discussion on the two most significant PCA peaks around $\sim$3~keV and $\sim$8~keV. The significance of these two peaks varies between  $\sim$1$-$4~$\sigma$  based on the errors derived from our Monte Carlo analysis.
We find significant variability of the energies and the relative strengths of these peaks in the PC(1,E) component, however,  the peaks do not all vary in the same manner between time intervals. 
For example, the PC(1,E) peak near the observed-frame energy of 3~keV, that is likely associated with \SXVI, does not vary in energy between time intervals $t_{2}$ and $t_{7}$, however, the PC(1,E) peak near the observed-frame energy of  8~keV (that is likely associated with \FeXXV/\FeXXVI) varies significantly in energy between $t_{2}$ and $t_{7}$. This may imply the presence of multiple components in the outflow with different outflow properties.  Another reason for the different behaviors of the PCA peaks between time intervals is that the ionization parameter and density of the outflow may be varying  leading to the presence of different atomic absorption lines in each time interval. 

The second largest outflow velocity of $v$ = 0.283 $\pm$ 0.006$c$ detected in time-interval $t_{7}$ was found to be significant at $>$~99.9\% confidence with Monte-Carlo simulations (Figure \ref{fig:f_test}) 
with the \FeXXV\ absorption line showing 3 data points significantly below the continuum and the \FeXXVI\ absorption line showing 4 data points significantly below the continuum (Figures  \ref{fig:res_time} and \ref{fig:t1356}). The largest ionization of $\log\xi (\rm erg~cm~s^{-1})$ =  5.15$_{-0.25}^{+0.20}$, inferred in time-interval $t_{1}$, was found for an absorption line detected to be significant at $\sim$ 99.8\% confidence with Monte-Carlo simulations (see Table 4). The two highest flux states (detected at $>$~99.9\% confidence) of $f_{3-10}$ = 5.79~$\times$ 10$^{-13}$~erg~s$^{-1}$~cm$^{-2}$ and  $f_{3-10}$ = 4.79 $\times$ 10$^{-13}$~erg~s$^{-1}$~cm$^{-2}$ were detected in time-intervals $t_{7}$ and $t_{1}$, respectively. We conclude that the relativistic outflow in \iras\ does not disappear at the highest flux and/or ionization levels, in contrast to what was concluded using background dominated spectra in previous studies (e.g., \cite{2017Natur.543...83P}; \cite{2018MNRAS.476.1021P}). On the contrary the weakest detection ($\sim$ 98\% confidence) of the Fe line is made in time-interval $t_{4}$ which has the lowest flux level of $f_{3-10}$ = 2.35~$\times$ 10$^{-13}$~erg~s$^{-1}$~cm$^{-2}$.

One important assumption when stacking spectra either by flux or by time is that the properties of the absorber  ($N_{H}$, $\xi$, $v_{\rm outflow}$, ionizing SED) and reflector have not varied significantly.  As shown in  Figures \ref{fig:bkg} and \ref{fig:lc}, the $r_{250}$ spectra stacked by flux show multiple absorption features in the 8 $-$ 11~keV range within a single stacked spectrum.  
On the other hand, as shown in Figures \ref{fig:res_time} and \ref{fig:t1356},  spectra stacked by time contain well defined Fe absorption lines and when fit with the  \verb+relxill+ + \verb+warmabs+ model provide acceptable fits in a statistical sense. As shown in Figure \ref{fig:lc}, the 3$-$10~keV flux of \iras\ changes by factors of up to $\sim$8 in time interval $t_{7}$ and by  $\sim$2  in 10~ks.
If the velocity of the outflow was tracking the short timescale variations of the 3$-$10~keV flux then one would expect to find multiple absorption lines smeared between 0.2$c$ $-$ 0.3$c$ in the stacked by time interval $r_{250}$ spectra. But this is not what we observe.  On the other hand, the $r_{250}$ spectra stacked by flux do show multiple components as expected for spectra stacked with different outflow properties.

\section{Equivalent Width of Fe Absorption Line Versus 3$-$10~keV Flux of \iras} \label{sec:width}

\begin{figure}[ht!]
\plotone{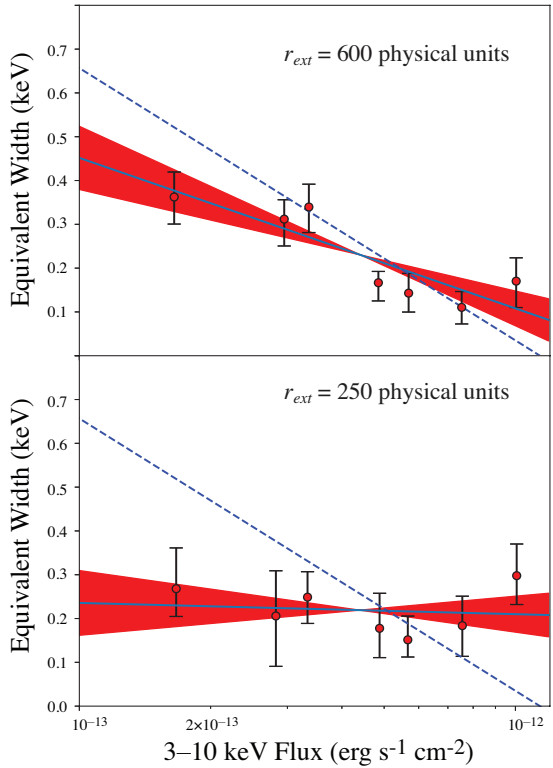}
\caption{Top: The equivalent width of the iron absorption line as a function of the 3$-$10~keV flux obtained from modeling the
stacked $r_{600}$ spectra of \iras. The stacking was based on the 0.3$-$10~keV count-rate ranges listed in Table \ref{tab:counts}.
We reproduce the anti-correlation found by \cite{2017MNRAS.469.1553P}. 
Bottom: The equivalent width versus 3$-$10~keV flux obtained from modeling the stacked $r_{250}$ spectra.
The stacking was based on the 0.3$-$10~keV count-rate ranges listed in Table \ref{tab:counts}. 
The dashed lines represent the best-fit found by \cite{2017MNRAS.469.1553P}. The solid lines represents the fits to the stacked $r_{250}$ and $r_{600}$ spectra
from this analysis. The shaded area represents the uncertainty of our fit to the $r_{250}$ data. The EW's obtained from the $r_{250}$ spectra are consistent with no correlation. \label{fig:ewc}}
\end{figure}

In  Figure \ref{fig:ewc} we show the equivalent width of the high-energy absorption lines as a function of the 3$-$10~keV flux obtained from modeling the stacked (based on their 0.3$-$10~keV count rates)  $r_{250}$ and $r_{600}$ spectra of \iras. 
Specifically, spectra with 0.3$-$10~keV count rates that lie within specific ranges listed in Table \ref{tab:counts} were stacked and the 3$-$10~keV flux of the each stacked spectrum was calculated. 
As we show in Figure \ref{fig:fluxranges}, there is significant scatter between the 0.3$-$10~keV count-rates and the 3$-$10~keV fluxes of \iras, however, we present these results to directly compare with results presented in \cite{2017MNRAS.469.1553P}.  We fit the equivalent width versus  3$-$10~keV flux data  with a straight line of slope $a$. 
For the fits obtained from modeling the stacked $r_{250}$ and $r_{600}$ spectra (based on their 0.3$-$10~keV count rates) we find slopes of $a$ = $-$0.03 $\pm$ 0.15 and $a$ = $-$0.35 $\pm$ 0.11, respectively.

\begin{figure}[ht!]
\plotone{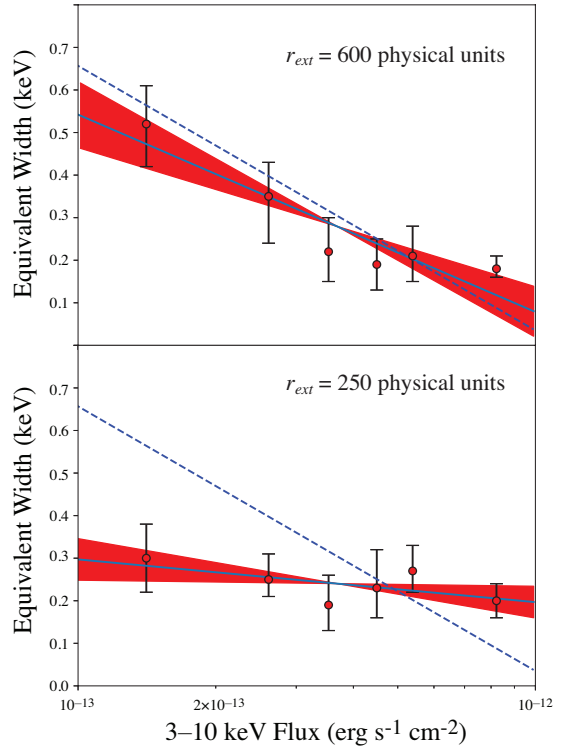}
\caption{Top: The equivalent width of the iron absorption line as a function of the 3$-$10~keV flux obtained from modeling the
stacked $r_{600}$ spectra of \iras. 
The stacking was based on the 3$-$10~keV fluxes shown in Figures \ref{fig:fluxranges} and \ref{fig:bkg}.
We reproduce the anti-correlation found by \cite{2017MNRAS.469.1553P}. 
Bottom: The equivalent width versus 3$-$10~keV flux obtained from modeling the stacked $r_{250}$ spectra. 
The stacking was based on the 3$-$10~keV fluxes shown in Figures \ref{fig:fluxranges} and \ref{fig:bkg}.
The dashed lines represent the best-fit found by \cite{2017MNRAS.469.1553P}. The solid lines represents the fits to the stacked $r_{250}$ and $r_{600}$ spectra
from this analysis. The shaded area represents the uncertainty of our fit to the $r_{250}$ data. The EW's obtained from the $r_{250}$ spectra are consistent with no correlation. \label{fig:ewf}}
\end{figure}

In Figure \ref{fig:ewf} we show the equivalent width of the high-energy absorption lines as a function of the 3$-$10~keV flux obtained from modeling the stacked (based on their 3$-$10~keV fluxes)  $r_{250}$ and $r_{600}$ spectra of \iras. The flux ranges used to stack spectra are shown in Figures \ref{fig:fluxranges} and \ref{fig:bkg}. 
For the fits to the equivalent width versus 3$-$10~keV data obtained from modeling the stacked $r_{250}$ and $r_{600}$ spectra (based on their 0.3$-$10~keV fluxes) we find slopes of $a$ = $-$0.1 $\pm$ 0.1 and $a$ = $-$0.46 $\pm$ 0.14, respectively.

We reproduce within errors the anti-correlation of the equivalent width versus 3$-$10~keV flux found by \cite{2017MNRAS.469.1553P} when performing a spectral analysis of the high-background stacked $r_{600}$ spectra of \iras\ both for stacking based on 0.3$-$10~keV count rates and for stacking based on 3$-$10~keV fluxes. 
More importantly, however, we find that when analyzing the stacked $r_{250}$ spectra of \iras, that are less affected by background than the  $r_{600}$ spectra, the anti-correlation disappears 
with the best-fit slopes being consistent with zero within errors.
We conclude that we find no significant dependence of the equivalent width of the high energy absorption lines with 3$-$10~keV flux in \iras.  
Another relevant point comes from Section 4 where we found that the energy and strength of the iron absorption lines are variable. 
Modeling spectra stacked according to their 3$-$10~keV fluxes is therefore likely to lead to unreliable and unrealistic results.   

\cite{2018MNRAS.476.1021P} performed simultaneous fits to the count-rate-selected full band EPIC-pn and RGS spectra of \iras\ and suggest (see their Figure 3) that the absorption lines weaken and seem to disappear at higher luminosities. We have demonstrated that the highly-ionized Fe absorption lines do not disappear at higher luminosities when background is reduced in the Epic-pn spectra. The Pinto et al. spectral results may have been affected in the simultaneous fits since their Epic-pn spectra are background dominated above $\sim$ 7~keV and the inferred properties of the outflowing ionized absorber for their full-band fits may be unreliable. The only significant soft X-ray features in their Epic-pn count-rate stacked spectra lie in the range of 0.5$-$1~keV. The origin of these features is uncertain and may be partly due to modeling the soft excess with a simple blackbody component. There are additional assumptions made in the Pinto et al. analysis that may have affected their spectral results.
For example, the line width, and the line-of-sight velocity of the outflowing absorber are fixed and tied in their spectral fits to the measured values of \cite{2017MNRAS.469.1553P}, whereas, our analysis shows that these quantities vary significantly. In addition, the spectra in the Pinto et al. analysis have been combined by count-rate over a $\sim$4~year period. Their analysis assumes that the properties of the outflowing absorber and accretion disk are the same for similar count-rate levels over a $\sim$4~year period.

\section{Energetics of the Relativistic Outflow of \iras} \label{sec:energetics}
Two important properties of an AGN's outflow that may indicate whether or not it is important in regulating the growth of the host galaxy are the mass-outflow rate ($\dot{M}$) and the ratio of the rate of kinetic energy ejected by the outflow to the AGN's bolometric luminosity ($\epsilon_{K}$). 

\begin{equation} \label{eq:energetics}
\epsilon_K=\frac{1}{2}\frac{\dot{M}v^2}{L_{bol}}, ~{\rm where}~
\dot{M}=4 \pi R^2 \rho v f_c=4 \pi f_c \frac{R^2}{\Delta R} N_{\rm
H} m_p v,
\end{equation}
where $f_c$, $N_{\rm H}$, $R$,  and $\Delta R$ are the covering fraction, column
density, radius, and thickness of the outflowing absorber, respectively.
We estimated $\dot{M}$ and $\epsilon_{K}$ for the two time intervals $t_5$ and $t_7$ for which 
a significant change in the outflow velocity is observed and the iron absorption lines are  detected at 
$>$ 99.9\%  confidence.

The global covering factor of the outflowing absorber is not constrained with the current observations. Specifically, there is no strong P-Cygni profile detected in any of the spectra of \iras. Several broad emission features detected between the observed energies of $\sim$ 7~keV  and $\sim$ 8~keV
are likely associated with the relativistic Fe line from the accretion disk. 
We assume a covering factor $f_c$ lying in the range of 0.3$-$0.6 based on the detected fraction of ultrafast outflows in Seyfert's (e.g., \cite{2010A&A...521A..57T}, \cite{2013MNRAS.430...60G}). To estimate the launching radius we assume that the maximum observed outflow velocity is produced by gas that has reached its terminal velocity, resulting in the approximation $R_{\rm launch}$ $\sim$ a few times $R_{s \rm}(c/v_{\rm wind} )^{2}$ , where $v_{wind}$ is the observed outflow velocity and $R_{\rm s}$ = $2GM/c^2$. 
For estimating the mass outflow rate and outflow efficiency, we assumed a fraction  $R/{\Delta}{R}$ ranging from 1 to 10 based on 
theoretical models of quasar outflows (e.g., \cite{2000ApJ...543..686P}).  We used a Monte Carlo approach to estimate the errors of $\dot{M}$ and $\epsilon_{\rm k}$. In Table \ref{tab:properties} we list the mass-outflow rate and the efficiency of the outflow for time intervals $t_{1}$, $t_{3}$, $t_{5}$, and $t_{7}$. The mass-outflow rates for time intervals $t_{5}$ and $t_{7}$  that are separated by $\sim$ 500~ks are $\dot{M}$~=~0.47$_{-0.31}^{+0.40}$~$M_{\odot}$~yr$^{-1}$ and  $\dot{M}$~=~0.20$_{-0.14}^{+0.22}$~$M_{\odot}$~yr$^{-1}$, respectively, which are comparable to the accretion rate of \iras\  which we estimate to be 1.8 $\times$ 10$^{-3}$($L_{44}$/$\eta$)$M_{\odot}$ ~yr$^{-1}$ $\sim$~0.6~$M_{\odot}$~yr$^{-1}$, where we assumed a typical accretion efficiency of $\eta$ = 0.1.
We estimate the outflow efficiencies for time intervals $t_{5}$ and $t_{7}$ to be 
$\epsilon_{\rm k}$~=~0.11$_{-0.07}^{+0.10}$ and $\epsilon_{\rm k}$~=~0.06$_{-0.04}^{+0.06}$, respectively. 
For comparison purposes we note that \cite{2017MNRAS.469.1553P} report a mass outflow rate of $\dot{M}$~=~$0.03{\Omega}M_{\odot}$~yr$^{-1}$, where $\Omega$ is the
solid angle of the wind. For their estimate, they assumed that the properties of the absorber did not vary over the $\sim$ one month observing period of \iras.
Specifically, they assumed for their estimate of $\dot{M}$ an ionization state of \FeXXV, a launching radius of 100~$r_{\rm g}$, a constant wind velocity of 0.244$c$, a column density of 8 $\times$ 10$^{22}$ cm$^{-2}$, and a ratio of the distance of the absorber from the black hole to the absorber thickness of $r/{\Delta}r$ = 1.
Our estimated mass-outflow rates differ from the \cite{2017MNRAS.469.1553P} reported value based on our revised analysis of the Epic-pn observations of \iras. 
Specifically, our estimates of the energetics of the wind of \iras\ are based on fitting photoionization models to the time-resolved $r_{250}$ spectra that show significant variability of the properties of the wind. Our mass-outflow rates listed in Table \ref{tab:properties} range between 0.1 $-$ 0.47$M_{\odot}$~yr$^{-1}$, assuming a covering fraction lying in the range of 0.3$-$0.6.

Our estimated ratios of the outflow mechanical luminosities to the bolometric luminosity, albeit with large uncertainties, are large enough, according to numerical simulations \citep{2005Natur.433..604D,2010MNRAS.401....7H,2012ApJ...745L..34Z}, for the outflow to unbind gas in the bulge of the host galaxy if this energy were efficiently transferred to the ISM.

\section{Conclusions} \label{sec:conclusions}

We have presented results from a re-analysis of {\sl XMM-Newton} observations of \iras\ to assess the influence of background on previously reported results. We also undertook a slightly different analysis approach in combining the spectra and investigated the change of the properties of the outflow as a function of 3$-$10~keV flux and time. The main conclusions of our spectral and timing analyses are the following.

\begin{enumerate}

\item{We confirm by applying PCA on the $r_{250}$ and $r_{600}$ spectra of \iras, taken from the total exposure time of $\sim$ 1.5~Ms spread over a month from 2016$-$07$-$08 to 2016$-$08$-$09,
the presence of significant variability of the spectrum of \iras\ at rest-frame energies of  1.33~keV, 1.81~keV, 2.51~keV, 3.20~keV, 4.40~keV, 5.38~keV, 6.06~keV, and 8.57~keV. These energies are similar to the ones reported in \cite{2017Natur.543...83P}. The significance of these individual peaks in the combined spectrum of \iras\ lies in the range of 1$-$4$\sigma$, as indicated by the thickness of the shaded regions of the Eigenvector PC(1,E) shown in Figure~\ref{fig:pca6} as derived from our Monte Carlo analysis.} The only difference is a slight increase of PC(1,E) for energies above observed-frame energies of 8~keV, when background contamination is reduced when using the $r_{250}$ spectra (see Figure~\ref{fig:pca}).

\item{We find that at energies above $\sim$~7~keV  background contamination is significant (see Figure \ref{fig:bkg}) for the source extraction regions used by \cite{2017MNRAS.469.1553P,2017Natur.543...83P}. 
When the background contamination in the spectral analysis is reduced by a factor of $\sim$8, by selecting smaller source extraction regions, we find several results that differ from the ones previously reported. Specifically, we do not find a constant velocity $v = 0.236c$~$\pm$~0.006$c$ outflow in \iras\ as reported in Parker et al, analyzing the same data.  Instead we find that the energy of the high-energy absorption line detected at $>$ 99.9\% confidence increases with 3$-$10~keV flux indicating an outflow with a velocity varying between $v \sim 0.2c$ and $v \sim 0.3c$. 
(see Figures \ref{fig:res_time}, \ref{fig:v_vs_flux}, and \ref{fig:unfolded})}

\item{PCA is a powerful tool in identifying spectral variability of the strengths of absorption lines. 
We applied time-resolved PCA to seven sequential time intervals of \iras\ to search for variability of the energies of the lines. We find significant variability of the energies and the relative strengths of the peaks of the PC(1,E) component, however, the peaks do not all vary in the same manner between time intervals (see Figure \ref{fig:pca6}). Possible reasons for the difference in the behavior of the PCA peaks are the presence of multiple components in the outflow and variability of the column density and/or ionization parameter of the outflowing absorber.} 

\item{We find that the equivalent width of the iron absorption line detected in \iras\ is not anti-correlated with 0.3$-$10~keV flux when background levels are reduced using the $r_{250}$ spectra. We find that the apparent anti-correlation is an artifact of background contamination.}


\item{ With the re-analysis of the \iras\ spectra we have discovered higher than previously reported velocity components of the outflow which is important in estimating the energetics of the outflow. In two time intervals labelled $t_5$ and $t_7$ (see Figure \ref{fig:lc}) of observations of \iras\ separated by about 500~ks  we estimate that the mass-outflow rate lies in the range of 
$\dot{M}$~=~0.47$_{-0.31}^{+0.40}$ $-$ 0.10$_{-0.07}^{+0.09}$  $M{\odot}$~yr$^{-1}$, which is comparable to the accretion rate of \iras.
In the same time intervals we find the fraction of kinetic to bolometric luminosity lies in the range of 
$\epsilon_{\rm k}$~=~0.11$_{-0.07}^{+0.10}$ $-$ 0.06$_{-0.04}^{+0.06}$, suggesting that this wind produces
 significant feedback to the host galaxy.}
\end{enumerate}

It is difficult to detect such high velocity outflowing components in nearby AGN due to the relative low effective area of current X-ray observatories at energies above 8~keV. 
In high redshift quasars, the iron absorption lines are redshifted to lower energies and are more easily detected which may partly explain the discovery of the highest outflow velocities in distant quasars.

\acknowledgments
We acknowledge financial support from NASA via the grants SAO GO6-17099X12 and NNX16AH33G.
We thank Massimo Cappi, Mauro Dadina, Cristian Vignali, Giorgio Lanzuisi, Margharita Giustini and Cristian Saez
for many useful comments and suggestions.  We greatly appreciate the useful comments made by the referee.


\begin{deluxetable*}{ccCrlcc}[b!]
\tablecaption{Log of Observations of \iras\ \label{tab:obslog}}
\tablecolumns{7}
\tablenum{1}
\tablewidth{0pt}
\tablehead{
\colhead{Date\tablenotemark{a}} &
\colhead{OBS ID} &
\colhead{Exposure} & \colhead{Net Exp\tablenotemark{b}} & \colhead{Net Count Rate\tablenotemark{c}} & \colhead{$f_{3-10}$\tablenotemark{d}} \\
\colhead{} & \colhead{} &
\colhead{(s)} & \colhead{(s)} &  \colhead{(s$^{-1}$)} & \colhead{}
}
\startdata
2016-08-09 & 0792180601 & 134,915  & 121,283 & 3.817 $\pm$ 0.006 & 7.86$^{+0.10}_{-0.12}$ \\
2016-08-07 & 0792180501 & 136,306 &  120,248 & 1.544 $\pm$ 0.004 & 4.19$^{+0.06}_{-0.10}$\\
2016-08-03 & 0792180401 & 139,717 &  124,604  & 3.810 $\pm$  0.006 &8.52$^{+0.08}_{-0.13}$ \\
2016-08-01 & 0792180301 & 139,417 &  120,431   & 0.715 $\pm$ 0.003 & 2.34$^{+0.06}_{-0.10}$ \\
2016-07-30 & 0792180201 & 139,414 &   125,938 &1.788 $\pm$  0.004  & 3.60$^{+0.05}_{-0.09}$ \\
2016-07-26 & 0792180101 & 139,919 &   127,435 & 1.518 $\pm$ 0.004 & 3.34$^{+0.07}_{-0.10}$ \\
2016-07-24 & 0780561701 & 139,715  &  125,831   &1.406 $\pm$  0.003  & 3.15$^{+0.07}_{-0.08}$ \\
2016-07-22 & 0780561601 & 139,711 &   127,405  & 2.880 $\pm$ 0.005 & 5.75$^{+0.08}_{-0.10}$ \\
2016-07-20 & 0780561501 & 139,716 &   127,743  & 1.220 $\pm$ 0.003 & 2.66$^{+0.07}_{-0.08}$ \\
2016-07-12 & 0780561401 & 137,015 &     80,172 & 2.221 $\pm$  0.005 & 4.82$^{+0.09}_{-0.09}$ \\
2016-07-10 & 0780561301 & 139,918  &  123,883  & 1.884  $\pm$ 0.004 & 3.92$^{+0.07}_{-0.09}$ \\
2016-07-08 & 0780560101 & 140,219   &   31,396  & 1.705 $\pm$ 0.007 & 3.27$^{+0.08}_{-0.25}$ \\
\enddata
\tablenotetext{a}{Date of exposure start.}
\tablenotetext{b}{Time is the effective exposure time remaining after the application of good time-interval tables and the removal of portions of the observation that were severely contaminated by background flaring.}
\tablenotetext{c}{Background-subtracted source counts including events with energies within the 0.4$-$10~keV band. The source counts and effective exposure times for the XMM-Newton observations refer to those obtained with the EPIC-pn instrument. The radii of the circular source and background extraction regions are 600~pu and 1600~pu, respectively.}
\tablenotetext{d}{The 3$-$10~keV flux of the $r_{600}$ spectra in units of 10$^{-13}$~erg~s$^{-1}$~cm$^{-2}$.}
\end{deluxetable*}

\begin{deluxetable*}{ccc}[b!]
\tablecaption{The count-rate regimes used in stacking the spectra of \iras\  \label{tab:counts}}
\tablecolumns{3}
\tablenum{2}
\tablewidth{0pt}
\tablehead{
\colhead{Count-Rate Regimes} &
\colhead{Count-Rate\tablenotemark{a}} &
\colhead{$f_{3-10}$\tablenotemark{b}}  \\
\colhead{} & \colhead{(s$^{-1}$)}  & \colhead{($\times$ 10$^{-13}$~erg~s$^{-1}$~cm$^{-2}$)}
}
\startdata
1 & 0 $-$ 0.6  & 1.65$^{+0.06}_{-0.09}$ \\
2 & 0.6 $-$ 0.95  & 2.95$^{+0.07}_{-0.10}$ \\
3 & 0.95 $-$ 1.4  & 3.36$^{+0.08}_{-0.18}$ \\
4 & 1.4 $-$ 1.95  & 4.85$^{+0.10}_{-0.12}$ \\
5 & 1.95 $-$ 2.3  & 5.75$^{+0.10}_{-0.27}$ \\
6 & 2.3 $-$ 3.2 & 7.53$^{+0.16}_{-0.18}$ \\
7 & $>$ 3.2 & 10.04$^{+0.22}_{-0.25}$ \\
\enddata
\tablenotetext{a}{The count-rates are calculated in the 0.3$-$10~keV observed-frame energy band of $r_{600}$ spectra of \iras. }
\tablenotetext{b}{$f_{3-10}$ are the 3$-$10~keV fluxes of the combined $r_{600}$ spectra of \iras\ of each count-rate regime.}
\end{deluxetable*}

\begin{deluxetable*}{ccCCrlcRRR}[b!]
\tablecaption{Properties of the Fe Absorption Line of  \iras\ as a Function of 3$-$10~keV Flux \label{tab:linepro_f}}
\tablecolumns{10}
\tablenum{3}
\tablewidth{0pt}
\tablehead{
\colhead{$f_{3-10}$~\tablenotemark{a}} &
\colhead{E$_{abs}$\tablenotemark{b}} &
\colhead{EW$_{abs}$\tablenotemark{c}} &
\colhead{N$_{abs}$\tablenotemark{d}} & \colhead{$\chi^{2}$~\tablenotemark{e}} & \colhead{$\nu$}\tablenotemark{f} & \colhead{$C$}\tablenotemark{g}  & \colhead{$F$}\tablenotemark{h} & \colhead{$P_{F}$\tablenotemark{i}}\\
\colhead{} & \colhead{(keV)} & \colhead{(keV)} &
\colhead{} & \colhead{} &  \colhead{} & \colhead{} & \colhead{} & \colhead{}
}
\startdata
1.29$^{+0.05}_{-0.08}$ & 8.61$_{-0.06}^{ +0.08}$  & -0.30$_{-0.08}^{+0.08}$   & $-$2.8$_{-0.8}^{+0.8}$ & 52.8(65.82) & 47(48)  & $>99\%$ &11.57 & 0.009\\
2.37$^{+0.07}_{-0.09}$ & 8.62$_{-0.10}^{+0.04}$   & -0.25$_{-0.06}^{+0.04}$ & $-$4.9$_{-1.6}^{+1.6}$  & 90.55(114.25) & 77(78) &  $>99\%$ & 20.16 & $<$0.001\\
2.37$^{+0.07}_{-0.09}$ & 9.67$_{-0.26}^{+0.05}$   & -0.34$_{-0.09}^{+0.07}$ & $-$5.0$_{-2.2}^{+2.2}$   & 104.08(114.25)  & 77(78) &$>99\%$& 7.53 & 0.01 \\
3.10$^{+0.08}_{-0.11}$ & 9.14$_{-0.10}^{+0.16}$   & -0.19$_{-0.07}^{+0.06}$ & $-$3.0$_{-1.8}^{+1.5}$ & 74.50(84.52)  &  82(84) & $99\%$ &5.51 & 0.01\\
4.23$^{+0.10}_{-0.14}$ &8.76$_{-0.07}^{+0.08}$    & -0.23$_{-0.09}^{+0.07}$  & $-$5.2$_{-2.9}^{+2.9}$ & 61.39(72.12) & 85(87)  & $>99\%$ &7.43 & 0.003\\
5.01$^{+0.13}_{-0.17}$ & 8.70$_{-0.04}^{+0.04}$   & -0.27$_{-0.06}^{+0.05}$ & $-$11.86$_{-2.8}^{+2.8}$  &  90.35(116.13) & 78(80) & $>99\%$& 11.13  & $<$ 0.001\\
7.65$^{+0.12}_{-0.14}$ & 9.38$_{-0.06}^{+0.09}$   & -0.20$_{-0.04}^{+0.04}$  & $-$8.4$_{-1.4}^{+1.4}$   &  149.90()167.60)   & 153(154)  & $>99\%$ &18.06 & $<$ 0.001 \\
7.65$^{+0.12}_{-0.14}$ &9.96$_{-0.10}^{+0.09}$    & -0.17$_{-0.05}^{+0.05}$ & $-$6.5$_{-1.5}^{+1.5}$  &   161.73(167.60)  & 153(154)  & $99\%$  & 5.55 & 0.015 \\
\enddata
\tablenotetext{a}{The 3$-$10~keV flux of the $r_{250}$ spectra (combined by 3$-$10~keV flux) listed in Figure \ref{fig:bkg} in units of 10$^{-13}$~erg~s$^{-1}$~cm$^{-2}$ }
\tablenotetext{b}{Rest-frame  energies of the blueshifted Fe absorption lines.}
\tablenotetext{c}{Rest-frame equivalent width of the blueshifted Fe absorption lines.}
\tablenotetext{d}{Normalization of the  blueshifted Fe absorption  line in units of 10$^{-7}$~photons~cm$^{-2}$~s$^{-1}$}
\tablenotetext{e}{$\chi^{2}$ of alternative model. The $\chi^{2}$ of the null model is quoted in the parenthesis.}
\tablenotetext{f}{Degrees of freedom of alternative mode. The degrees of freedom of the null model is quoted in the parenthesis. The null model includes a simple absorbed power-law and the alternative model includes one or two Gaussian absorption lines.}
\tablenotetext{g}{$\chi^{2}$ confidence detection levels of the blueshifted Fe absorption lines.} 
\tablenotetext{h}{$F$-statistic between the null and alternative model. }
\tablenotetext{i}{The probability of exceeding this $F$ value as determined from the Monte Carlo simulations.}
\tablecomments{These properties were obtained from fits to the $r_{250}$ spectra of \iras\ combined in the 3$-$10~keV flux ranges indicated in Figure \ref{fig:bkg}. }
\end{deluxetable*}

\begin{deluxetable*}{cccCRRRRRRR}[b!]
\tablecaption{Properties of the Fe Absorption Line of  \iras\ as a Function of Time \label{tab:linepro_t}}
\tablecolumns{11}
\tablenum{4}
\tablewidth{0pt}
\tablehead{
\colhead{Time Interval\tablenotemark{a}} &
\colhead{$f_{3-10}$\tablenotemark{b}} &
\colhead{E$_{abs}$\tablenotemark{c}} &
\colhead{EW$_{abs}$\tablenotemark{d}} &
\colhead{N$_{abs}$\tablenotemark{e}} & \colhead{$\chi^{2}~$\tablenotemark{f}} & \colhead{$\nu$}\tablenotemark{g} & \colhead{$C$}\tablenotemark{h}  & \colhead{$F$}\tablenotemark{i} & \colhead{$P_{F}$~\tablenotemark{j}}\\
\colhead{} & \colhead{(keV)} & \colhead{(keV)} &
\colhead{} & \colhead{} &  \colhead{} & \colhead{} & \colhead{} & \colhead{}
}
\startdata
1 &4.1$_{-0.2}^{+0.1}$ & 8.88$_{-0.05}^{ +0.05}$  &-0.30$_{-0.06}^{+0.07}$  & $-$7$_{-2}^{ +2}$  & 22.67(37.32) & 37(39)  & $>99\%$ &11.96 & 0.002\\
2 &4.6$_{-0.3}^{+0.2}$ & 9.36$_{-0.41}^{+0.08}$  &-0.41$_{-0.16}^{+0.17}$  & $-$9$_{-4}^{ +4}$ &  18.62(25.20) & 21(23) &  $>90\%$ & 3.70 &  0.06\\
3 &3.3$_{-0.1}^{+0.1}$ & 8.80$_{-0.08}^{+0.13}$ &-0.35$_{-0.11}^{+0.08}$  & $-$8$_{-2}^{+2}$ &   94.77(128.22)  & 95(97) &$>99\%$& 16.77 & $<$0.001 \\
4 &2.4$_{-0.1}^{+0.1}$ & 8.53$_{-0.17}^{+0.32}$   & -0.17$_{-0.10}^{+0.12}$ &$-$3$_{-1}^{ +1}$ &  98.6(110.25)   &  77(79) & $99\%$ &4.55 &  0.02\\
5 &2.9$_{-0.1}^{+0.1}$ & 8.66$_{-0.03}^{+0.04}$   & -0.27$_{-0.05}^{+0.04}$  & $-$5$_{-1}^{ +1}$ &   124.85(160.56) & 95(97)  & $>99\%$ &13.59 & $<$0.001\\
6 &8.7$_{-0.2}^{+0.1}$  & 9.45$_{-0.25}^{+0.25}$  &-0.44$_{-0.14}^{+0.10}$  & $-$23$_{-11}^{ +11}$ &   76.9(91.7) & 68(71) & $99\%$& 4.4  & 0.01\\
7 &5.8$_{-0.1}^{+0.1}$ & 9.36$_{-0.04}^{+0.06}$  &-0.29$_{-0.08}^{+0.06}$  & $-$5$_{-1}^{ +1}$ &   161.89(183.55) & 156(158) & $>99\%$& 10.44  & $<$ 0.001\\
\enddata
\tablenotetext{a}{The time intervals used to extract the spectra of \iras\ are shown in Figure \ref{fig:lc}.}
\tablenotetext{b}{The 3$-$10~keV flux of the $r_{250}$ spectra (combined by 3$-$10~keV flux) listed in Figure \ref{fig:bkg} in units of $\times$ 10$^{-13}$~erg~s$^{-1}$~cm$^{-2}$}
\tablenotetext{c}{Rest-frame energy of the blueshifted Fe absorption lines.}
\tablenotetext{d}{Rest-frame equivalent width of the blueshifted Fe absorption lines.}
\tablenotetext{e}{Normalization of the  blueshifted Fe absorption  line in units of  $\times$ 10$^{-7}$~photons~cm$^{-2}$~s$^{-1}$}
\tablenotetext{f}{$\chi^{2}$ of alternative model. The $\chi^{2}$ of the null model is quoted in the parenthesis. The null model includes a simple absorbed power-law and the alternative model includes one or two Gaussian absorption lines.}
\tablenotetext{g}{Degrees of freedom of alternative mode. The degrees of freedom of the null model is quoted in the parenthesis.}
\tablenotetext{h}{$\chi^{2}$ confidence detection levels of the blueshifted Fe absorption lines.} 
\tablenotetext{i}{$F$-statistic between the null and alternative model.}
\tablenotetext{j}{The probability of exceeding this $F$ value as determined from the Monte Carlo simulations.}
\tablecomments{These properties were obtained from fits to the $r_{250}$ spectra of \iras\ extracted from the intervals indicated in Figure \ref{fig:lc}. }
\end{deluxetable*}

\begin{deluxetable*}{ccCrlcc}[b!]
\tablecaption{Properties of the Outflow of \iras\ \label{tab:properties}}
\tablecolumns{7}
\tablenum{5}
\tablewidth{0pt}
\tablehead{
\colhead{Time Interval\tablenotemark{a}} &
\colhead{N$_{\rm H}$\tablenotemark{b}} &
\colhead{$\log\xi$} & \colhead{$v_{\rm abs}$} & \colhead{$\dot{M}$} & \colhead{$\epsilon_{\rm k}$\tablenotemark{c}} \\
\colhead{} & \colhead{$\times$ 10$^{24}$ cm$^{-2}$ } &
\colhead{(erg~cm~s$^{-1}$)} & \colhead{$(c)$} &  \colhead{(M$_{\odot}~yr^{-1}$)} & \colhead{}
}
\startdata
$t1$ & 0.74$_{-0.35}^{+0.58}$ & 5.15$_{-0.25}^{+0.20}$  & 0.226$_{-0.007}^{+0.005}$ &  0.45$_{-0.30}^{+0.43}$  & 0.19$_{-0.12}^{+0.18}$   \\
$t3$ & 0.62$_{-0.26}^{+0.17}$ & 4.5$_{-0.25}^{+0.20}$  & 0.194$_{-0.005}^{+0.005}$ &  0.32$_{-0.19}^{+0.23}$  & 0.10$_{-0.07}^{+0.06}$   \\
$t5$ & 1.0$_{-0.6}^{+1.5}$ & 3.7$_{-0.1}^{+0.1}$  & 0.173$_{-0.006}^{+0.005}$ &  0.47$_{-0.31}^{+0.40}$  & 0.11$_{-0.07}^{+0.10}$   \\
$t7$ & 0.13$_{-0.06}^{+0.09}$ & 4.3$_{-0.3}^{+0.3}$  & 0.283$_{-0.006}^{+0.006}$ & 0.10$_{-0.07}^{+0.09}$ & 0.06$_{-0.04}^{+0.06}$   \\
\enddata
\tablenotetext{a}{Time intervals listed in Figure \ref{fig:lc}}
\tablenotetext{b}{The best-fit value of the column density of the outflowing absorber obtained by fitting the $r_{250}$ spectra of \iras\ with the warmabs photoionization model.}
\tablenotetext{c}{The efficiency of the outflow, $\epsilon_{\rm k}$, is defined as the ratio of the outflow mechanical luminosity to the bolometric luminosity of \iras.}
\tablecomments{The properties of the outflow were determined by fitting a model that consists of a power-law modified by Galactic absorption, an outflowing intrinsic ionized absorber and relativistically blurred X-ray reflection to the $r_{250}$ spectra of \iras. }
\end{deluxetable*}



\end{document}